\documentclass[prd,amsmath,amssymb,longbibliography,superscriptaddress,twocolumn,nofootinbib,10pt]{revtex4-1}

\pdfoutput=1

\usepackage{graphicx}
\usepackage{dcolumn}
\usepackage{bm}
\usepackage{amssymb}
\usepackage{latexsym}
\usepackage{booktabs}
\usepackage{amsmath, mathrsfs, bm}
\usepackage{multirow}
\usepackage{threeparttable}

\usepackage[colorlinks=true, linkcolor=blue, citecolor=blue]{hyperref}

\begin{document}


\title{Detecting stochastic gravitational wave background from cosmic strings with next-generation detector networks: Component separation based on a multi-source astrophysical foreground noise model}

\author{Geng-Chen Wang}
\affiliation{Liaoning Key Laboratory of Cosmology and Astrophysics, College of Sciences, Northeastern University, Shenyang 110819, China}
\author{Hong-Bo Jin}
\affiliation{Institute for Frontiers in Astronomy and Astrophysics, Beijing Normal University, Beijing, China}
\affiliation{National Astronomical Observatories, Chinese Academy of Sciences, Beijing 100101, China}
\affiliation{The International Centre for Theoretical Physics Asia-Pacific, University of Chinese Academy of Sciences, Beijing 100190, China}
\author{Xin Zhang}\thanks{zhangxin@mail.neu.edu.cn (corresponding author)}
\affiliation{Liaoning Key Laboratory of Cosmology and Astrophysics, College of Sciences, Northeastern University, Shenyang 110819, China}
\affiliation{MOE Key Laboratory of Data Analytics and Optimization for Smart Industry, Northeastern University, Shenyang 110819, China}
\affiliation{National Frontiers Science Center for Industrial Intelligence and Systems Optimization, Northeastern University, Shenyang 110819, China}

\begin{abstract}

Detecting stochastic gravitational wave background (SGWB) from cosmic strings is crucial for unveiling the evolutionary laws of the early universe and validating non-standard cosmological models. In light of extensive existing studies on the detection capabilities of third-generation gravitational wave detectors for cosmic string signals, this work provides a systematic evaluation focused specifically on the performance of next-generation ground-based detector networks. By constructing a hybrid signal model that incorporates multi-source astrophysical foregrounds, including compact binary coalescences (CBCs) and compact binary hyperbolic encounters (CBHEs), we develop a parameter estimation methodology based on multi-component signal separation.
Numerical simulations using one-year observational data reveal three key findings: (1) The CE4020ET network, comprising the Einstein Telescope (ET-10 km) and the Cosmic Explorer (CE-40 km and CE-20 km), achieves nearly one order of magnitude improvement in constraining the cosmic string tension $G\mu$ compared to individual detectors, reaching a relative uncertainty $\Delta G\mu / G\mu < 0.5$ for $G\mu > 3.5 \times 10^{-15}$ under standard cosmological framework; (2) The network demonstrates enhanced parameter resolution in non-standard cosmological scenarios, providing a novel approach to probe pre-Big Bang Nucleosynthesis cosmic evolution; (3) Enhanced detector sensitivity amplifies CBHE foreground interference in parameter estimation, while precise modeling of such signals could further refine $G\mu$ constraints by $1-2$ orders of magnitude. This research not only quantifies the detection potential of third-generation detector networks for cosmic string models but also elucidates the intrinsic connection between foreground modeling precision and cosmological parameter estimation accuracy, offering theoretical foundations for optimizing scientific objectives of next-generation gravitational wave observatories.

\end{abstract}
\maketitle

\section{Introduction}\label{sec1}
The detection of gravitational waves (GWs) from astrophysical sources by the LIGO-Virgo collaboration in 2015~\cite{LIGOScientific:2017adf} marked a significant milestone in GW astronomy. As of the latest observations, 90 binary black holes (BBHs), 2 neutron star-black holes (NSBHs) and 2 binary neutron stars (BNSs) have been detected~\cite{Nitz:2021zwj}. In addition to these prominent and individually resolvable GW events, a large number of signals from various sources remain too weak to be detected individually. The incoherent combination of these signals forms a stochastic gravitational wave background (SGWB)~\cite{Allen:1997ad, Sathyaprakash:2009xs, Caprini:2018mtu, Christensen:2018iqi, Renzini:2022alw}. 

SGWBs originate from various sources, which can be broadly classified into cosmological and astrophysical sources. 
Cosmological sources include cosmic strings (CSs)~\cite{Nielsen:1973cs, Kibble:1976sj, Vilenkin:1984ib, Witten:1985fp, Hindmarsh:1994re, Polchinski:2004ia}, first-order phase transitions~\cite{Krauss:1991qu, Kosowsky:1991ua, Kosowsky:1992rz, Kosowsky:1992vn, Kamionkowski:1993fg, Hindmarsh:2013xza}, inflation~\cite{grishchuk1975amplification, starobinsky1979relict, starobinskii1979spectrum, Rubakov:1982df, Grishchuk:1993te, Barnaby:2011qe}, and other early universe phenomena.
Astrophysical sources include compact-binary coalescences (CBCs) (including BBHs, NSBHs, BNSs and white dwarf binaries)~\cite{Meacher:2015iua, Farmer:2003pa}, compact-binary hyperbolic encounters (CBHEs)~\cite{Capozziello:2008ra, DeVittori:2012da, Garcia-Bellido:2017knh, Garcia-Bellido:2017qal, Garcia-Bellido:2021jlq}, stellar core collapse~\cite{Ferrari:1998ut, Buonanno:2004tp, Crocker:2015taa, Crocker:2017agi, Finkel:2021zgf} and others.
These SGWB signals are expected to carry crucial information about cosmic structure formation and evolutionary processes, offering unprecedented insights into diverse physical phenomena ranging from compact object dynamics to early universe evolution.

The detection and analysis of SGWBs, especially those originating from cosmological processes, has profound scientific significance. 
CSs are one-dimensional topological defects that form through spontaneous symmetry breaking during early-universe phase transitions and are predicted to exist throughout cosmic history~\cite{Kibble:1976sj}. As the CS network evolves, it produces closed loops that decay, emitting GWs. The cumulative uncorrelated contributions from these emissions form a SGWB~\cite{Vilenkin:1981bx, Vachaspati:1984gt, Damour:2000wa, Damour:2004kw, Siemens:2006yp, Olmez:2010bi}.
According to the Standard Model of cosmology, the universe experienced inflation, followed by radiation domination, then transitioned to a matter-dominated phase, and finally entered an accelerated expansion driven by dark energy. However, current observations offer limited insights into the era before Big Bang Nucleosynthesis (BBN)~\cite{Boyle:2005se, Boyle:2007zx}, and the actual cosmic evolution can significantly differ from the standard scenario. Non-standard cosmologies may significantly affect the SGWB spectrum from CSs, particularly the high-frequency part~\cite{Freese:1995vp, Jeannerot:2003qv, Rocher:2004my, Cui:2017ufi, Cui:2018rwi, Guedes:2018afo, Gouttenoire:2019kij, Allahverdi:2020bys, Ghoshal:2023sfa}. Through the observation of CSs, we anticipate establishing significant constraints on cosmological models.

Detecting the SGWB from cosmic strings poses significant challenges, as the corresponding gravitational wave signals are inherently weak and expected to be obscured by intense astrophysical foreground noise dominated by compact binary coalescences (CBCs). Second-generation detectors, including Advanced LIGO~\cite{Harry:2010zz, LIGOScientific:2014pky}, Advanced Virgo~\cite{VIRGO:2014yos}, and KAGRA~\cite{Aso:2013eba}, have not yet confirmed the detection of any cosmological SGWB signals, with current results providing only upper limits. This is primarily attributed to their limited sensitivity and the present lack of effective methodologies to sufficiently resolve and subtract the predicted CBC foreground contamination~\cite{KAGRA:2021duu, KAGRA:2021kbb}.
With the deployment of next-generation (NG) observatories, we expect to observe a more comprehensive SGWB of cosmological origin. The NG observatory network is expected to detect hundreds of thousands of CBC events annually~\cite{Evans:2021gyd, Borhanian:2022czq, Ronchini:2022gwk, Iacovelli:2022bbs, Evans:2023euw}, as well as many CBHE events~\cite{Mukherjee:2020hnm}. These observations will enable the development of more precise astrophysical foreground models, facilitating the detection of the SGWB from CSs.

Building upon existing research on cosmic string detection with third-generation gravitational wave detectors~\cite{Martinovic:2020hru, Branchesi:2023mws, Meijer:2023yhn}, this study further investigates the joint detection capabilities of NG ground-based gravitational wave detector networks for CSs. Our objective is to systematically evaluate the networks' ability to detect non-standard features in the SGWB spectrum generated by cosmic strings under conditions involving more complex astrophysical foreground noise, thereby providing new avenues for reconstructing the evolutionary history of the universe prior to BBN.
We employ the component separation method~\cite{Parida:2015fma} to isolate CS signals from astrophysical foregrounds. Our analysis includes both the standard cosmological model for CSs and the impact of non-standard cosmologies on their SGWB spectrum. We consider a more comprehensive range of astrophysical foreground noise, including both CBC and CBHE foregrounds. Additionally, we investigate the impact of excluding the CBHE foreground on the precision of CS constraints.

This work is organized as follows. In Sec.~\ref{sec2}, we discuss the sensitivity of NG ground-based GW detector networks. In Sec.~\ref{sec3}, we describe the SGWB from CSs, CBCs, and CBHEs. In Sec.~\ref{sec4}, we introduce the component separation method and data simulation techniques, and simulate one year of observational data. In Sec.\ref{sec5}, we present the results of CS observations derived from the one-year simulated data after removing astrophysical foregrounds. The conclusion is given in Sec.~\ref{sec6}. We work with natural units $c = \hbar = 1$.

\section{Sensitivity for detector networks}\label{sec2}
\begin{figure*}
    \centering
    \includegraphics[width=1\linewidth]{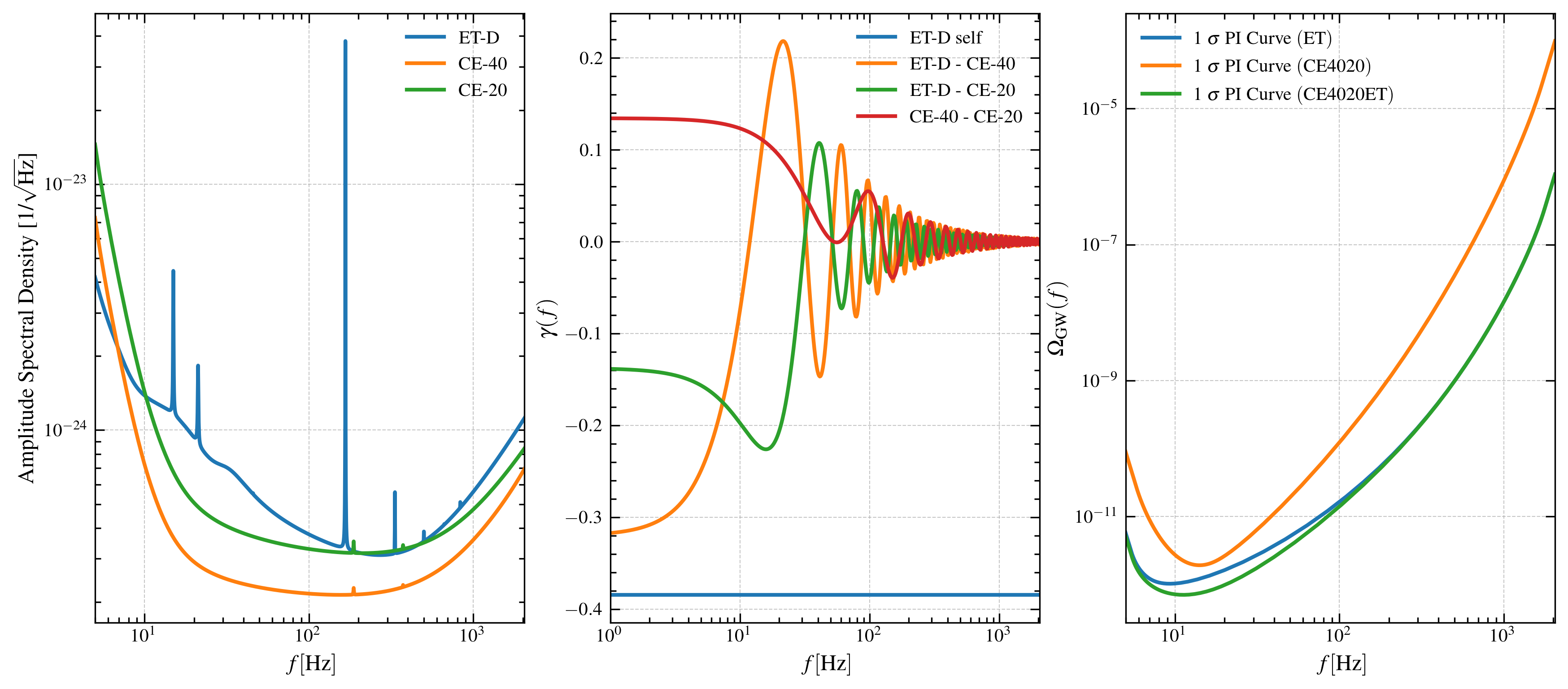}
    \caption{Sensitivity of different detector networks in detecting the SGWB. \textit{Left}: Amplitude spectral densities for the three detectors in the networks. \textit{Middle}: Overlap reduction functions for each baseline in different detector networks, normalized such that $\gamma(f) = 1$ for co-located and co-aligned L-shaped detectors. \textit{Right}: $1 \sigma$ PI curves showing the sensitivity of the various detector networks considered in this study to the stochastic gravitational-wave background after one year of observation~\cite{Thrane:2013oya}.}
    \label{ASDPIorf}
\end{figure*}

We focus on NG ground-based GW detectors, the Einstein Telescope (ET)~\cite{Punturo:2010zz} and Cosmic Explorer (CE)~\cite{Reitze:2019iox}, which are expected to be operational around 2035. For ET sensitivity, we adopt the baseline configuration, ET-D, which consists of a triangular array of three Michelson interferometers, each with 10-km-long arms~\cite{Hild:2010id}. For CE sensitivity, we consider two equal-arm L-shaped interferometers, CE-40 and CE-20, with arm lengths of 40 km and 20 km, respectively~\cite{Evans:2023euw}. The strain sensitivities of these detectors are shown in the left panel of Fig.~\ref{ASDPIorf}. We consider three observational scenarios for NG ground-based GW detectors: 1. A single ET detector, which includes three Michelson interferometers with 10 km arm lengths. 2. The CE network (CE4020), which consists of two NG detectors: CE-40 with a 40 km arm length and CE-20 with a 20 km arm length. 3. The CE and ET network (CE4020ET), which includes three NG detectors: CE-40 with a 40 km arm length, CE-20 with a 20 km arm length, and ET-D with 10 km arm lengths arranged in a triangular configuration. 

The sensitivity of a detector network depends not only on the one-sided strain power spectral density (PSD) of individual detectors but also on their relative positions and orientations, described by the overlap reduction function $\gamma(f)$. The overlap reduction function $\gamma(f)$ quantifies the sky- and polarization-averaged cross-correlation between the responses of two detectors as follows~\cite{Allen:1997ad}:
\begin{align}
    \gamma(f)=\frac{5}{8 \pi} \int  {\rm d}\hat{\mathbf{n}} F^A_1(f, \hat{\mathbf{n}}) F^A_2(f, \hat{\mathbf{n}}) e^{- i 2 \pi f \hat{\mathbf{n}} \Delta \mathbf{x}}. 
\end{align}
Here, $\hat{\mathbf{n}}$ denotes the unit vector pointing to a direction on the sky, $\Delta \mathbf{x}$ represents the separation vector between detectors 1 and 2, and $A$ indicates the polarization basis. Using Table II of Ref.~\cite{Gupta:2023lga}, we determined the positions and orientations of the CE and ET observatories. The middle panel of Fig.~\ref{ASDPIorf} illustrates the overlap reduction functions for different detector networks.

In this study, we choose the frequency range from 5 Hz to 512 Hz. This lower cut-off frequency is chosen to align with the proposed design sensitivities of CE and ET~\cite{fritschel2022report}.
The upper cut-off frequency is set at 512 Hz, following standard analysis, since the overlap reduction function of this detector pair has negligible power beyond this frequency. To assess the ability of different detector networks to detect SGWB signals over a one-year observation period, we use the $1 \sigma$ power-law integrated (PI) curves described in Ref.~\cite{Thrane:2013oya}, shown in the right panel of Fig.~\ref{ASDPIorf}.

\section{Components of the Stochastic Gravitational Wave Background}\label{sec3}
The SGWB is characterized by its spectral emission, which is the primary target of stochastic GW searches. The spectrum is commonly parametrized by the GW fractional energy density spectrum $\Omega_{\rm GW}(f)$, defined as 
\begin{align}
    \Omega_{\rm GW}(f) = \frac{1}{\rho_c} \frac{{\rm d} \rho_{\rm GW}(f)}{{\rm d}\ln f},
\end{align}
where ${\rm d} \rho_{\rm GW}(f)$ represents the energy density of GWs from the source within the frequency range $f$ to $f + {\rm d}f$, and $\rho_c$ denotes the critical energy density of the universe. 

The fractional energy density spectra of SGWB from different sources exhibit unique characteristics. This paper considers three sources of SGWB: CSs, CBCs, and CBHEs. Among these, the SGWB generated by CSs is the primary signal of interest, while the other two are treated as foreground noise.

\subsection{Stochastic gravitational wave background from cosmic strings}
{This work investigates the gravitational wave background generated by networks of local cosmic strings. Local cosmic strings arise from the spontaneous breaking of a local gauge symmetry and represent cosmologically significant topological defects predicted by well-motivated particle physics models~\cite{Kibble:1976sj}, such as supersymmetric grand unified theories (SUSY GUTs). It is important to note that cosmic strings may also originate from more complex symmetry-breaking scenarios~\cite{Jeannerot:2003qv} or manifest as fundamental strings in superstring theory. In contrast, global cosmic strings, formed through the breaking of global symmetries, exhibit distinctly different phenomenological features, including their gravitational coupling and radiation efficiency, compared to local cosmic strings~\cite{Baeza-Ballesteros:2023say, Klaer:2017qhr, Chang:2019mza}.
The primary source of GWs from a cosmic string network is the continuous formation of loops through network fragmentation, which occurs when strings self-intersect or collide. These loops oscillate relativistically and act as persistent sources of GWs throughout cosmic history, collectively generating a SGWB through the superposition of numerous uncorrelated emissions~\cite{Vilenkin:1981bx, Vachaspati:1984gt, Damour:2000wa, Damour:2004kw, Siemens:2006yp, Olmez:2010bi}.
The impact of cosmology on the GW spectrum stems from two competing effects: a red-tilt due to the redshifting of GW energy density, and a blue-tilt from the loop production rate. During radiation domination, these effects cancel, resulting in a nearly scale-invariant (flat) spectrum~\cite{Gouttenoire:2019kij}. Since the frequency band of ground-based GW detectors probes the part of the CS-generated SGWB originating from the radiation-dominated era, characterizing this spectrum offers a powerful probe into the expansion history of the universe.}

CSs have been extensively studied in standard cosmology~\cite{Kibble:1984hp, Caldwell:1991jj, Polchinski:2006ee, DePies:2007bm, Siemens:2006yp, Olmez:2010bi, Kuroyanagi:2012jf, Kuroyanagi:2012wm, Sanidas:2012ee, Sousa:2013aaa, Blanco-Pillado:2013qja, Blanco-Pillado:2017oxo, Blanco-Pillado:2017rnf, Ringeval:2017eww, LIGOScientific:2017ikf, Jenkins:2018nty, Auclair:2019wcv, Blanco-Pillado:2019tbi, LIGOScientific:2021nrg, LISACosmologyWorkingGroup:2022jok}.
Additionally, several CS models in non-standard cosmology have been proposed~\cite{Freese:1995vp, Jeannerot:2003qv, Rocher:2004my, Cui:2017ufi, Cui:2018rwi, Guedes:2018afo, Gouttenoire:2019kij, Allahverdi:2020bys, Ghoshal:2023sfa}. These models modify the CS GW spectrum. Specifically, the high-frequency spectrum no longer remains flat but exhibits different spectral indices.
Next, we will discuss the CS GW spectrum in both standard cosmological and non-standard cosmological models.

\subsubsection{Standard cosmological model}
We adopt the template from Ref.~\cite{Sousa:2020sxs} to represent the GWs of CSs. This template accurately predicts the amplitude of the radiation-era plateau and fits well with the high-frequency cutoff and low-frequency peak from loops decaying during the matter era, providing a detailed description of the SGWB produced by CSs. In this template, we constrain only one parameter: the string tension $G\mu$. The other parameters, such as the loop size $\alpha$ and the total power of CS emission $\Gamma$, are fixed at $\alpha = 0.1$ and $\Gamma = 50$. Based on Planck 2018 data~\cite{Planck:2018vyg}, we set $H_0 = 100~h\rm~km/s/Mpc$, with $h = 0.678$, and assume a flat universe where radiation and matter density parameters at the present time are $\Omega_m = 0.308$ and $\Omega_r = 9.1476 \times 10^{-5}$, respectively. GW emission from CS loops arises from three distinct periods: loops formed and decayed during the radiation period, loops formed during the radiation period and decayed during the matter period, and loops formed during the matter period. 

For loops formed and decayed in the radiation era, the SGWB takes the following form~\cite{Sousa:2020sxs}:
\begin{widetext}
    \begin{align}
    	\Omega_{\mathrm{CS}}^r(f) = \frac{128}{9} \pi A_r \Omega_r \frac{G \mu}{\epsilon_r} \left[\left(\frac{f(1 + \epsilon_r)}{B_r \Omega_m / \Omega_r + f} \right)^{3 / 2} - 1 \right],\label{3}
    \end{align}
\end{widetext}
where $\epsilon_r = \alpha \xi_r / \Gamma G \mu$, $\xi_r = 0.271$, $A_r = 0.54$, and $B_r = 4 H_0 \Omega_r^{1 / 2} / (\Gamma G \mu)$. For loops formed during the radiation period and decayed during the matter period, the SGWB is given by the following expression~\cite{Sousa:2020sxs}:
\begin{widetext}
	\begin{align}
		\Omega_{\mathrm{CS}}^{rm}(f) = & \, 32 \sqrt{3} \pi (\Omega_{m} \Omega_{r})^{3 / 4} H_{0} \frac{A_{r}}{\Gamma} \frac{(\epsilon_{r} + 1)^{3 / 2}}{f^{1 / 2} \epsilon_{r}} \nonumber \\
		& \times \left\{\frac{\left({\Omega_m} / {\Omega_r} \right)^{1 / 4}}{\left(B_m \left({\Omega_m} / {\Omega_r} \right)^{1 / 2} + f \right)^{1 / 2}} \left[2 + \frac{f}{B_m \left({\Omega_m} / {\Omega_r} \right)^{1 / 2} + f} \right] - \frac{1}{(B_m + f)^{1 / 2}} \left[2 + \frac{f}{B_m + f} \right] \right\},
	\end{align}
\end{widetext}
where $B_m=3H_0\Omega_m^{1/2}/(\Gamma G\mu)$. For loops formed during the matter period, the SGWB is expressed as follows~\cite{Sousa:2020sxs}:
\begin{widetext}
	\begin{align}
	\Omega_{\mathrm{CS}}^m(f)=54 \pi H_0 \Omega_m^{3 / 2} \frac{A_m} \Gamma \frac{\epsilon_m + 1}{\epsilon_m} \frac{B_m} f \Bigl\{\frac{2 B_m + f}{B_m(B_m + f)} - \frac1f \frac{2 \epsilon_m + 1}{\epsilon_m(\epsilon_m + 1)} + \frac2f \log \Bigl(\frac{\epsilon_m + 1}{\epsilon_m}\frac{B_m}{B_m + f} \Bigr) \Bigr\},\label{5}
	\end{align}
\end{widetext}
where $\epsilon_m = \epsilon_r \xi_m / \xi_r$. We have the parameters $\xi_m = 0.625$ and $A_m = 0.039$.

Therefore, the SGWB generated by CSs can be well approximated as
\begin{align}
    \Omega_{\rm CS}(f)=\Omega_{\rm CS}^r(f)+\Omega_{\rm CS}^{rm}(f)+\Omega_{\rm CS}^m(f), \label{Omega_CS}
\end{align}
for $\alpha \geq \Gamma G\mu$. Since this model excludes the effective number of relativistic degrees of freedom (DoF) in the Standard Model, the CS GW spectrum in the high-frequency range can be approximated as a constant.

\subsubsection{Non-standard cosmological model}
The modification of the CS GW spectrum under non-standard cosmology is discussed in Ref.~\cite{Blanco-Pillado:2024aca}. It is assumed that before the recent radiation era, at temperatures $T > T_{\rm rd}$, the energy density of the universe is dominated by a new energy component, whose energy density scales as $\rho \propto a^{-3(1 + w)}$, where $w$ is the equation-of-state parameter for this period. In this case, for high frequencies $f > f_{\rm rd}$, the CS gravitational
wave spectrum $\Omega^\ast_{\rm CS}(f)$ is modified as follows:
\begin{align}
    \Omega_{\rm CS}^\ast(f) = &
    \begin{cases}
        \Omega_{\rm CS}(f)  & \text{for } f \leq f_{\rm rd}, \\ 
        \Omega_{\rm CS}(f_{\rm rd}) \left(\frac{f}{f_{\rm rd}} \right)^{-d} & \text{for } f > f_{\rm rd},\label{8}
    \end{cases} \\
    \text{where } d = &
    \begin{cases}
        1 & \text{for } w \leq \frac{1}{9}, \\
        -2 \frac{3w - 1}{3w + 1} & \text{for } w > \frac{1}{9}. \label{d}
    \end{cases}
\end{align}
The characteristic frequency $f_{\rm rd}$, above which the spectrum is modified, is related to the temperature of the universe at the onset of the recent radiation era, $T_{\rm rd}$, through the following analytical approximation~\cite{Cui:2018rwi}:
\begin{align}
    f_{\rm rd} = \left(8.67 \times 10^{-3} {\rm Hz} \right) \left(\frac{T_{\rm rd}}{\rm GeV} \right) \left(\frac{10^{-12}}{\alpha G \mu} \right)^{1/2}, \label{f_rd}
\end{align}
in this case, we do not consider the effect of the effective relativistic DoF and set $T_{\rm rd} = 1 \rm GeV$.
For values of $w$ ranging from $w = 1/9$ to $w = 1$ , the spectrum evolves differently.
For $w \leq 1/9$, the predicted spectrum remains unchanged, closely matching the spectrum of strings formed during inflation~\cite{Guedes:2018afo}.
For $1/9 < w < 1/3$, the spectrum exhibits a negative slope for $f \geq f_{\rm rd}$.
When $w = 1/3$, the spectrum corresponds to that of the standard cosmological model.
For $w > 1 / 3$, the spectrum exhibits a positive slope for $f \geq f_{\rm rd}$. 
When $w = 1$, the results align with the spectrum of CSs during the kination-dominated era immediately following inflation~\cite{Gouttenoire:2019kij}.
Additionally, Ref.~\cite{Zhu:2023gmx} analyzes a combination of pulsar timing array (PTA), BBN, and cosmic microwave background data, yielding $w = 0.44^{+0.52}_{-0.40}$ at the 95\% confidence level.
Therefore, we set $w = 1$, $w = 1/2$, $w = 1/3$, and $w = 1/4$ to investigate these different scenarios.

\begin{figure}
    \centering
    \includegraphics[width=8.6cm]{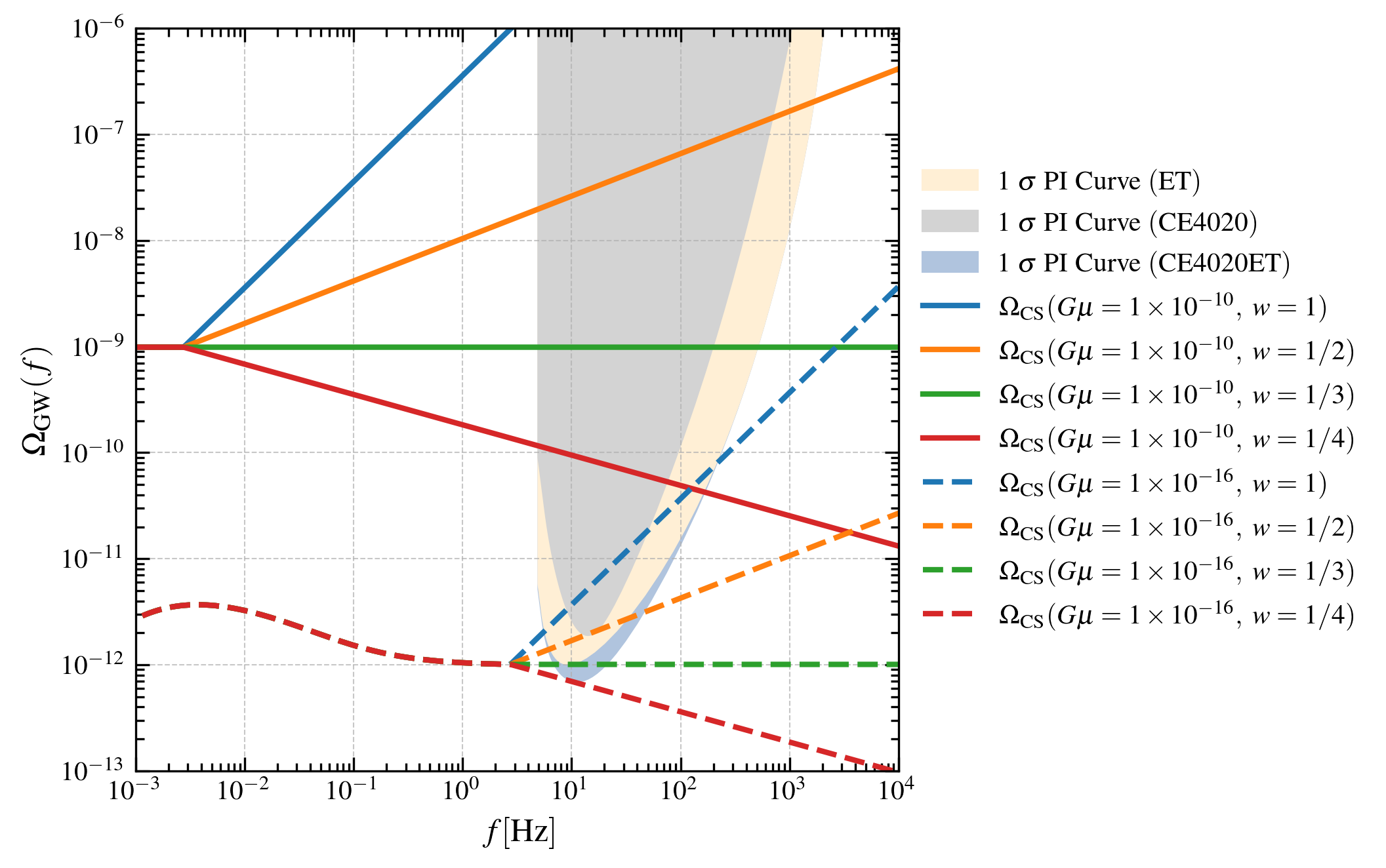}
    \caption{GW fractional energy density spectrum in modified pre-BBN scenarios. This figure illustrates examples of the GW fractional energy density spectrum generated by CSs in modified pre-BBN scenarios with a reheating temperature of $T_{\rm rd} = 1 \rm GeV$. The shaded area represents the $1 \sigma$ PI sensitivity window of three networks~\cite{Thrane:2013oya}. The solid line represents the case with $G\mu = 1 \times 10^{-10}$, and the dashed line represents the case with $G\mu = 1 \times 10^{-16}$. The lines in different colors represent different values of $w$. For frequencies greater than 5 Hz, the spectrum can be approximated by a power-law function.}  
    \label{Omega_CS_ast}
\end{figure}

In Fig.~\ref{Omega_CS_ast}, we present examples of the GW fractional energy density spectrum generated by CSs in modified pre-BBN scenarios with $T_{\rm rd} = 1$ GeV. We note that for $T_{\rm rd} = 1 \rm GeV$ and $G\mu \geq 1 \times 10^{-16}$, $f_{\rm rd}$ is always less than 5 Hz, which represents the lower cut-off frequency we have set. Therefore, we express the power-law function as:
\begin{align}
    \Omega_{\rm CS}(f) = \Omega_{\rm CS} \left(\frac{f}{f_{\ast}} \right)^{\alpha_{\rm CS}},
\end{align}
where $f_{\ast} = 25 \rm Hz$, $\Omega_{\rm CS} = \Omega_{\rm CS}(f_\ast)$ as defined in Eq.~(\ref{Omega_CS}), and $\alpha_{\rm CS} = -d$ as defined in Eq.~(\ref{d}).
{We note that our calculations in Eqs.~(3)--(5), (8) and (9) rely on a simplified approach: we assume cosmic string loops emit gravitational waves mainly at their fundamental frequency. While the complete theory includes all harmonic modes (following an $n^{-4/3}$ spectrum), we only consider the fundamental mode in this work. This simplification is reasonable because including higher harmonics has been shown to have little effect on the high-frequency part of the gravitational wave background~\cite{Blanco-Pillado:2017oxo,Sousa:2020sxs}. Our method is consistent with previous studies that use similar simplifications to make the analysis tractable~\cite{Blanco-Pillado:2024aca}. }

\subsection{Stochastic gravitational wave background from compact-binary coalescences}
In actual detections, the data include both GW signals from CSs and astrophysical foreground noise. In the sensitive frequency range of ground-based GW detectors, astrophysical foreground noise primarily originates from the CBCs GW background. Proposed NG ground-based detectors are expected to detect a broad range of astrophysical CBC sources, but they face challenges from subtraction residues when removing resolvable CBC signals $\Omega_{\rm CBC, res}(f)$, which must be minimized to reduce their impact on cosmological SGWB searches~\cite{Wu:2011ac, Zhu:2012xw, Regimbau:2016ike, Sharma:2020btq, Regimbau:2022mdu}. Beyond $\Omega_{\rm CBC, res}(f)$, the incoherent superposition of unresolved CBC signals leads to an astrophysical SGWB, known as $\Omega_{\rm CBC, unres}(f)$~\cite{Sachdev:2020bkk, Zhou:2022nmt, Zhou:2022otw, Zhong:2022ylh, Bellie:2023jlq, Zhong:2024dss}. Therefore, effectively,
\begin{align}
    \Omega_{\rm CBC}(f) = \Omega_{\rm CBC, res}(f) + \Omega_{\rm CBC, unres}(f).
\end{align}
Based on the GWTC-3 results from the third observing run (O3) of Advanced LIGO and Advanced Virgo~\cite{KAGRA:2021duu}, we consider the $\Omega_{\rm CBC, res}$, which includes contributions from BNS, NSBH, and BBH mergers, given by: $\Omega_{\rm BNS}(25 \, \rm Hz) = 0.6^{+1.7}_{-0.5} \times 10^{-10}$, $\Omega_{\rm NSBH}(25 \, \rm Hz) = 0.9^{+2.2}_{-0.7} \times 10^{-10}$, and $\Omega_{\rm BBH}(25 \, \rm Hz) = 5.0^{+1.4}_{-1.8} \times 10^{-10}$. We also use the result from Table 2 in Ref.~\cite{Bellie:2023jlq}, where $\Omega_{\rm CBC, unres}(25 \, \rm Hz) = 1.0^{+1.0}_{-0.6} \times 10^{-10}$.

For simplicity, we modeled the observational outcomes using a power-law function for the CBC foregrounds, contributed by both $\Omega_{\rm CBC, res}(f)$ and $\Omega_{\rm CBC, unres}(f)$:
\begin{align}
    \Omega_{\rm CBC}(f) = \Omega_{\rm CBC} \left(\frac{f}{f_\ast} \right)^{\alpha_{\rm CBC}},
\end{align}
where $f_\ast = 25 \rm Hz$, $\Omega_{\rm CBC} = \Omega_{\rm BNS}(f_\ast) + \Omega_{\rm NSBH}(f_\ast) + \Omega_{\rm BBH}(f_\ast) + \Omega_{\rm CBC, unres}(f_\ast) = 7.50 \times 10^{-10}$, and fixing the spectral index parameter to $\alpha_{\rm CBC} = 2/3$~\cite{Marassi:2011si}.

\subsection{Stochastic gravitational wave background from compact-binary hyperbolic encounters}
With the deployment of NG ground-based GW detectors, we are poised to observe a more complete SGWB of astrophysical origin~\cite{Mukherjee:2020hnm}. Here, we consider an additional important source of SGWB arising from overlapping GW bursts caused by CBHEs~\cite{Capozziello:2008ra, DeVittori:2012da, Garcia-Bellido:2017knh, Garcia-Bellido:2017qal, Garcia-Bellido:2021jlq}. When considering two interacting black holes (BHs), it is possible that they may not form bound systems, depending on the initial conditions, but instead produce single scattering events. Unlike CBCs, where the signal can last for many cycles, in the case of hyperbolic encounters, the signal is more appropriately viewed as a transient~\cite{Morras:2021atg}.
It is often argued that bound compact clusters, such as globular clusters (GCs), may be one of the dominant channels for hyperbolic encounters in the universe~\cite{dymnikova1982bursts, Kocsis:2006hq, OLeary:2008myb}. Therefore, encounters in GCs are expected to provide the sufficient number of events for an SGWB. This rate depends on the number of BHs in a cluster and the number of clusters in a galaxy. 

We refer to Ref.~\cite{Kerachian:2023gsa} to calculate the dimensionless GW energy density spectrum for CBHEs, $\Omega_{\rm CBHE}(f)$: 
\begin{eqnarray}
    \Omega_{\rm CBHE}(f) = \frac{n_{\rm gc}}{V\rho_{\rm c}}\int^{\infty}_{0}\frac{G(z)\mathcal{K}(z)dz}{(1+z)^3H(z)},
\end{eqnarray}
where $n_{\rm gc}$ is the number of GCs per Milky Way-equivalent galaxy (MWEG), $V$ is the comoving volume up to redshift $z$, $G(z)$ is the number of MWEGs between redshift $z$ and $z + {\rm d}z$ as given in Eq.~(32) of Ref.~\cite{Kerachian:2023gsa}, $\mathcal{K}(z)$ represents the probability per unit time of encounters occurring within each GC as given in Eq.~(36) of Ref.~\cite{Kerachian:2023gsa}, $\rho_{\rm c}$ is the energy density of the universe, and $H(z)$ is the Hubble parameter at redshift $z$.

We considered the most optimistic scenario from Table 1 of Ref.~\cite{Kerachian:2023gsa} and expressed the most optimistic scenario in the form of a power-law spectrum.
\begin{eqnarray}
    \Omega_{\rm CBHE}(f)=\Omega_{\rm CBHE}\left(\frac{f}{f_\ast}\right)^{\alpha_{\rm CBHE}},
\end{eqnarray}
where $f_\ast = 25\rm Hz$, $\Omega_{\rm CBHE} = 6.84 \times 10^{-12}$, and $\alpha_{\rm CBHE} = 9/5$.

\begin{figure}
    \centering
    \includegraphics[width=8.6cm]{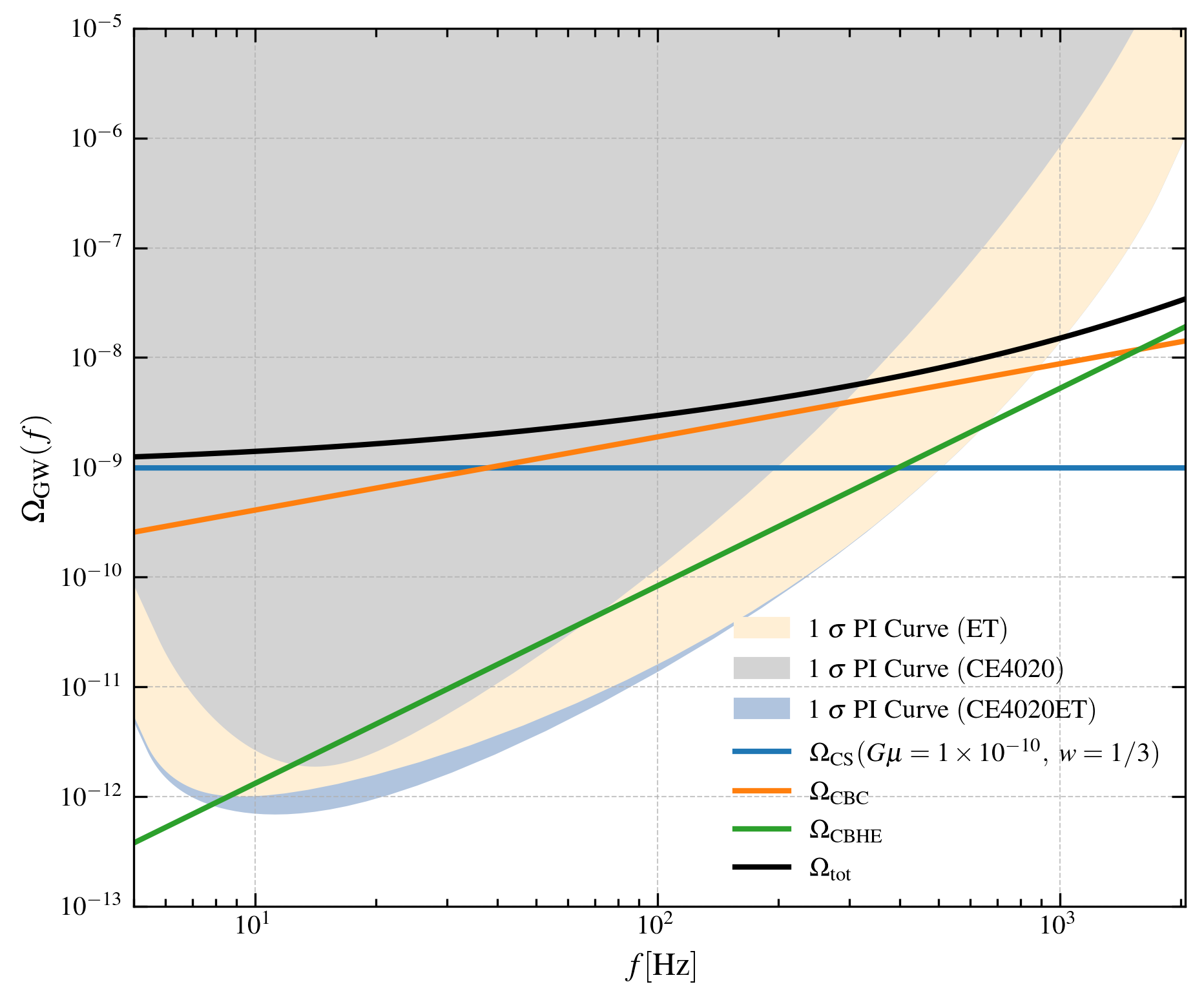}
    \caption{GW fractional energy density spectrum from multiple sources. The shaded area represents the $1 \sigma$ PI sensitivity window of three networks~\cite{Thrane:2013oya}. The blue line depicts the GW generated by CSs ($G\mu = 1 \times 10^{-10}$, $w = 1/3$), the orange line depicts the GW from CBCs, the green line depicts the GW from CBHEs, and the black line represents the total energy spectrum of GWs generated by the three sources mentioned above.}
    \label{PI}
\end{figure}

Therefore, the total energy spectrum related to GWs discussed in this paper is as follows
\begin{eqnarray}
    \Omega_{\rm GW}(f) = \Omega_{\rm CS}(f) + \Omega_{\rm CBC}(f) + \Omega_{\rm CBHE}(f).
\end{eqnarray}
The SGWB spectrum for the GW sources mentioned above is shown in Fig.~\ref{PI}, including contributions from CSs, CBCs, and CBHEs.

\section{Method}\label{sec4}
In this section, we use the component separation method from Ref.~\cite{Parida:2015fma} to evaluate the ability of three networks to estimate $G\mu$ under different data scenarios. 

\subsection{Component separation}
The total observed signals $s_1(t)$ and $s_2(t)$ for a single baseline in a network, consisting of two detectors (1 and 2), can be expressed as its respective (discrete) short-term Fourier transforms (SFTs) over time segments of duration $\Delta T$:
\begin{align}
    \tilde{s}_{1, t}(f) = \tilde{h}_{1, t}(f) + \tilde{n}_{1, t}(f),\\
    \tilde{s}_{2, t}(f) = \tilde{h}_{2, t}(f) + \tilde{n}_{2, t}(f),
\end{align}
where $\tilde{h}_{1, t}(f)$ and $\tilde{h}_{2, t}(f)$ represents the SFT of the strain signal, and $\tilde{n}_{1, t}(f)$ and $\tilde{n}_{2, t}(f)$ represent the SFTs of the uncorrelated random noise in each detector. Here, $t$ is a time-stamp marking a segment. In the small signal limit, where the variance of the strain signal is much smaller than the variance of the noise, the following expression is~\cite{Parida:2015fma}:
\begin{align}
    \widetilde{s}_{1, t}^\ast(f) \widetilde{s}_{2, t^{\prime}} \left(f^{\prime} \right) \approx 
    & \delta_{t t^{\prime}} \delta_{f f^{\prime}} \frac{\Delta T}{2} \frac{3 H_0^2}{10 \pi^2} \gamma(f) f^{-3} \Omega_{\mathrm{GW}}(f) \nonumber \\
    & +\widetilde{n}_{1, t}^\ast(f) \widetilde{n}_{2, t^{\prime}} \left(f^{\prime} \right),
\end{align}
where $\delta_{ij}$ is the Kronecker delta, which equals 1 when $i = j$ and 0 otherwise.

If the background consists of multiple components, and component $\alpha$ has a known spectral shape $\mathcal{F}^\alpha = (f/f_\ast)^\alpha$ with an amplitude of $\Omega_\alpha$, then, 
\begin{align}
    \Omega_{\rm GW} = \sum_\alpha  \Omega_\alpha \mathcal{F}^\alpha(f),
\end{align}
we apply the same method to express the convolution equation~\cite{Parida:2015fma},
\begin{align}
    \mathbf{C} = \mathbf{K}\cdot\mathbf{\Omega} + \mathbf{N},
    \label{convolution}
\end{align}
where
\begin{align}
    \mathbf{\Omega} & \equiv \Omega_\alpha, \\
    \mathbf{C} & \equiv \widetilde{s}_{1, t}^\ast(f) \widetilde{s}_{2, t}(f), \\
    \mathbf{N} & \equiv \widetilde{n}_{1, t}^\ast(f) \widetilde{n}_{2, t}(f), \\
    \mathbf{K} & \equiv \frac{\Delta T}{2} \frac{3 H_0^2}{10 \pi^2} \gamma(f) f^{-3} \mathcal{F}^\alpha(f).
\end{align}
If the one-sided noise PSDs of the data segments from the two detectors are denoted by $P_{1, t}(f)$ and $P_{2, t}(f)$, respectively, the covariance of $\mathbf{N}$ is given by~\cite{Parida:2015fma}
\begin{align}
    \mathcal{N} \equiv \left\langle \mathbf{N}^* \mathbf{N} \right\rangle = \delta_{t t^{\prime}} \delta_{f f^{\prime}} \left(\frac{\Delta T}{2} \right)^2 P_{1, t}(f) P_{2, t^{\prime}}(f^{\prime}).
\end{align}

The convolution equation Eq.~(\ref{convolution}) has a standard Maximum Likelihood solution for $\mathbf{\Omega}$, which is given by~\cite{Parida:2015fma}
\begin{align}
    \hat{\mathbf{\Omega}} = \mathbf{\Gamma}^{-1} \cdot \mathbf{X},
\end{align}
where
\begin{align}
    \mathbf{X} & = \mathbf{K}^\dagger \cdot \mathcal{N}^{-1} \cdot \mathbf{C},\\
    \mathbf{\Gamma} & = \mathbf{K}^\dagger \cdot \mathcal{N}^{-1} \cdot \mathbf{K}.
\end{align}
$\hat{\mathbf{\Omega}}$ is an unbiased estimator of the amplitudes of $\mathbf{\Omega}$. $\mathbf{\Gamma}$ is the Fisher information matrix for the estimated values, and its inverse is the noise covariance matrix, denoted by $\mathbf{\Sigma} = \mathbf{\Gamma}^{-1}$.

Based on the derivation in Ref.~\cite{Parida:2015fma}, the final set of formulas for the joint estimation of the components and their variance from real data are as follows:
\begin{align}
    \hat{\Omega}_\alpha & = \Gamma^{-1}_{\alpha\beta} X_\alpha =  \sum_\beta[B^{-1}]_{\alpha\beta} Y_\beta, \label{Omega} \\
    \Sigma_{\alpha\beta} & = \Gamma^{-1} = \left(\frac{10 \pi^2}{3 H_0^2} \right)^2 [M^{-1}]_{\alpha \beta}. \label{Sigma}
\end{align}
For convenience in calculation, we define the following~\cite{Parida:2015fma}:
\begin{align}
    M_{\alpha \beta} & = \sum_{I,t,f} \left|\gamma_I(f) \right|^2 \frac{\mathcal{F}^\alpha(f) \mathcal{F}^\beta(f)}{f^6 P_{I, 1, t}(f) P_{I, 2, t}(f)}, \label{M} \\
    Y_\alpha & = \Gamma^{-1}_{\alpha\alpha} X_\alpha = \left(\frac{10 \pi^2}{3 H_0^2} \right) \frac{2}{\Delta T} \frac{1}{M_{\alpha \alpha}} \nonumber \\ 
    & \times \sum_{I,t,f} \gamma_I(f) \frac{\mathcal{F}^\alpha(f) \widetilde{s}_{I, 1, t}^\ast(f) \widetilde{s}_{I, 2, t}(f)}{f^3 P_{I, 1, t}(f) P_{I, 2, t}(f)}, \label{Y} \\
    B_{\alpha \beta} & = \frac{\Gamma_{\alpha \beta}}{\Gamma_{\alpha \alpha}} = \frac{M_{\alpha \beta}}{M_{\alpha \alpha}}. \label{B}
\end{align}
Here, the index $I$ refers to a single baseline (pair of detectors) in a network.

\subsection{Simulation data}
To systematically investigate the impact of precise astrophysical foreground modeling on the detection capability of cosmic string signals, we designed three data scenarios in each detector. The key distinction among them lies in whether the signal components injected into the simulated data match the template model components used in Bayesian parameter estimation. The specific configurations are summarized below: \\
1. The injected signal model is given by $\Omega_{\rm GW}(f) = \Omega_{\rm CBC}(f) + \Omega_{\rm CBHE}(f)$, and the analysis template is chosen to be identical to the injection model. This configuration serves as a null test to validate the robustness of our methodology, specifically examining whether it can correctly infer the absence of cosmic string signals when only astrophysical foregrounds are present.\\
2. The injected signal model is given by $\Omega_{\rm GW}(f) = \Omega_{\rm CS}(f) + \Omega_{\rm CBC}(f) + \Omega_{\rm CBHE}(f)$, and the analysis template intentionally excludes the CBHE component. This configuration investigates the impact of incomplete foreground modeling by deliberately omitting the CBHE component during signal separation, allowing quantitative assessment of how proper inclusion of this foreground improves cosmic string detection sensitivity and parameter estimation accuracy.\\
3. The injected signal model is given by $\Omega_{\rm GW}(f) = \Omega_{\rm CS}(f) + \Omega_{\rm CBC}(f) + \Omega_{\rm CBHE}(f)$, and the analysis template is chosen to be identical to the injection model. This configuration demonstrates the optimal constraining power on cosmic string parameters under a complete and precise modeling framework.\\

\begin{figure}
    \centering
    \includegraphics[width=1\linewidth]{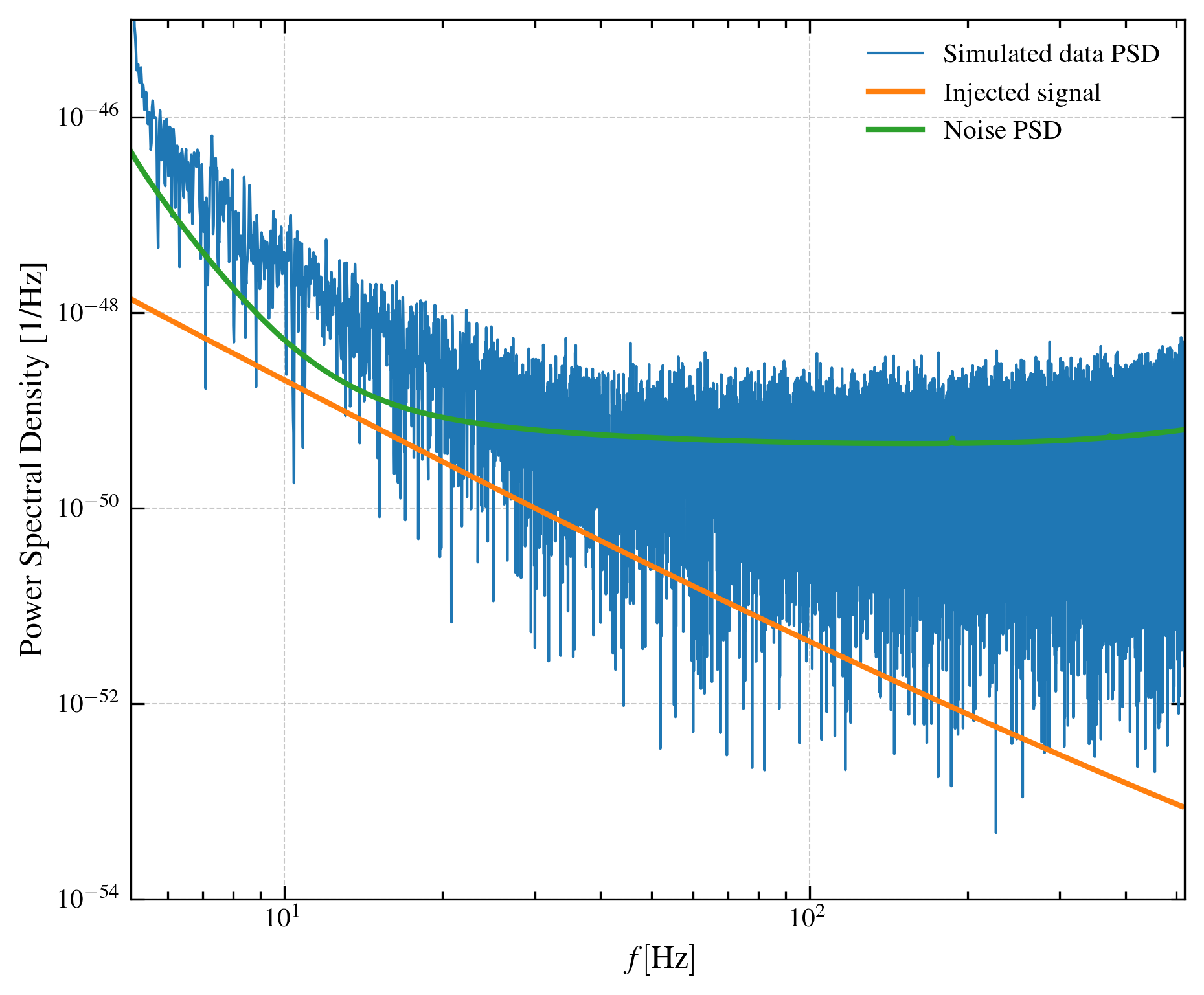}
    \caption{An example of simulated data generated by CE-40. The injected signal (orange) with $G\mu = 1 \times 10^{-10}$ and noise PSDs  (green) are plotted together with the calculated PSD (blue) of a simulated data segment duration of 16 s.}
    \label{Simulated}
\end{figure}

We use \texttt{pygwb}~\cite{Renzini:2023qtj} to simulate one year of time-series data at a sampling rate of 1024 Hz for each detectors and perform spectral analysis with a frequency resolution of 1/8 Hz and a segment duration of 16 s, with 50\% overlap between segments. An example of the simulated data is presented in Fig.~\ref{Simulated}. For each baseline in three networks, we calculate the one-sided noise PSD $P_{1, t}(f)$ and $P_{2, t}(f)$, as well as the one-sided cross PSD $C_{12, t}(f)$.
\begin{align}
    P_{1, t}(f) & = \frac{2}{\Delta T} |\tilde{n}_{1, t}(f)|^2, \\
    P_{2, t}(f) & = \frac{2}{\Delta T} |\tilde{n}_{2, t}(f)|^2, \\
    C_{12, t}(f)& = \frac{2}{\Delta T}  \tilde{s}_{1, t}^\ast(f) \tilde{s}_{2, t}(f).
\end{align}

\section{Results}\label{sec5}

\begin{figure}
    \centering
    \includegraphics[width=1\linewidth]{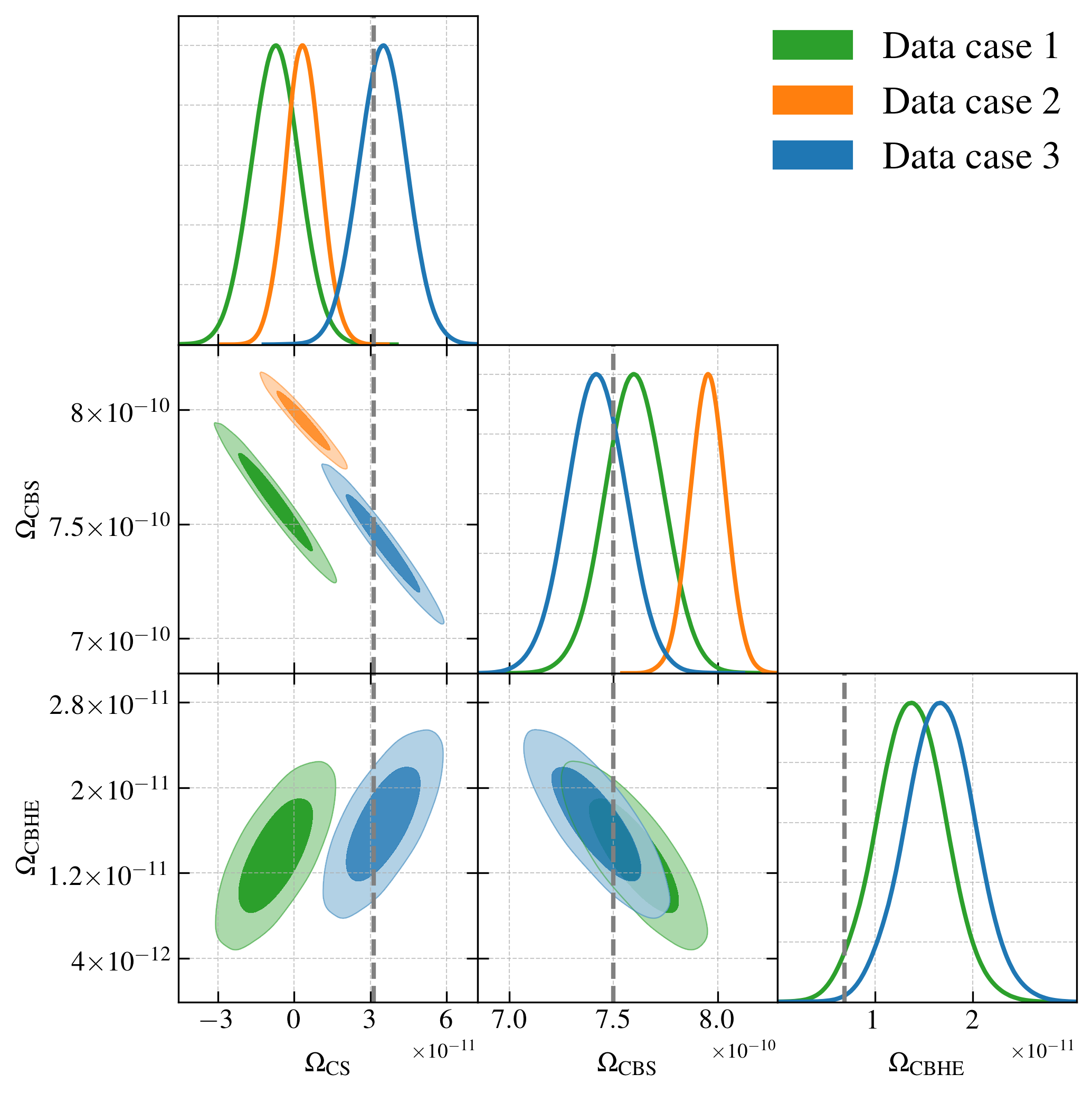}
    \caption{Marginalized distributions and confidence contours for GW component separation in CE4020ET network. For data case 1, 2, and 3 with $G\mu = 1 \times 10^{-13}$, $w = 1/3$ in the CE4020ET network, the one-dimensional marginalized distributions of individual components and the two-dimensional marginalized contours represent the 68.3\% and 95.4\% confidence levels, respectively. The green color represents the result of simulated data containing only astrophysical foreground noise. The blue color represents the component separation results with the full consideration of both CBC and CBHE foregrounds, while the orange color represents the results without considering the contribution of the CBHE foreground during component separation. The gray dashed line marks the injected true value. Using the component separation method, the amplitudes of multiple components are effectively estimated.}
    \label{CS+CBC+CBHE}
\end{figure}

In this section, we simulate one year of observational data to calculate the one-sided noise PSDs $P_{1, t}(f)$ and $P_{2, t}(f)$, as well as the one-sided cross PSD $C_{12, t}(f)$. This allows us to express Eqs.~(\ref{M} -- \ref{B}) and compute Eq.~(\ref{Omega}) and (\ref{Sigma}). The results of the joint estimation of the amplitudes of components with different spectral shapes are shown in Fig.~\ref{CS+CBC+CBHE}. In data case 1, the separation results for simulated data containing only astrophysical foreground noise suggest that the estimated amplitude of the CS component is negligible. In data cases 2 and 3, we analyze simulated data consisting of three components: CS signals, as well as CBC and CBHE foreground noise. The results show that the estimation of the CS signal in data case 3 is more accurate, highlighting the significance of incorporating the CBHE foreground.

\begin{figure*} 
    \centering
    \includegraphics[width=1\linewidth]{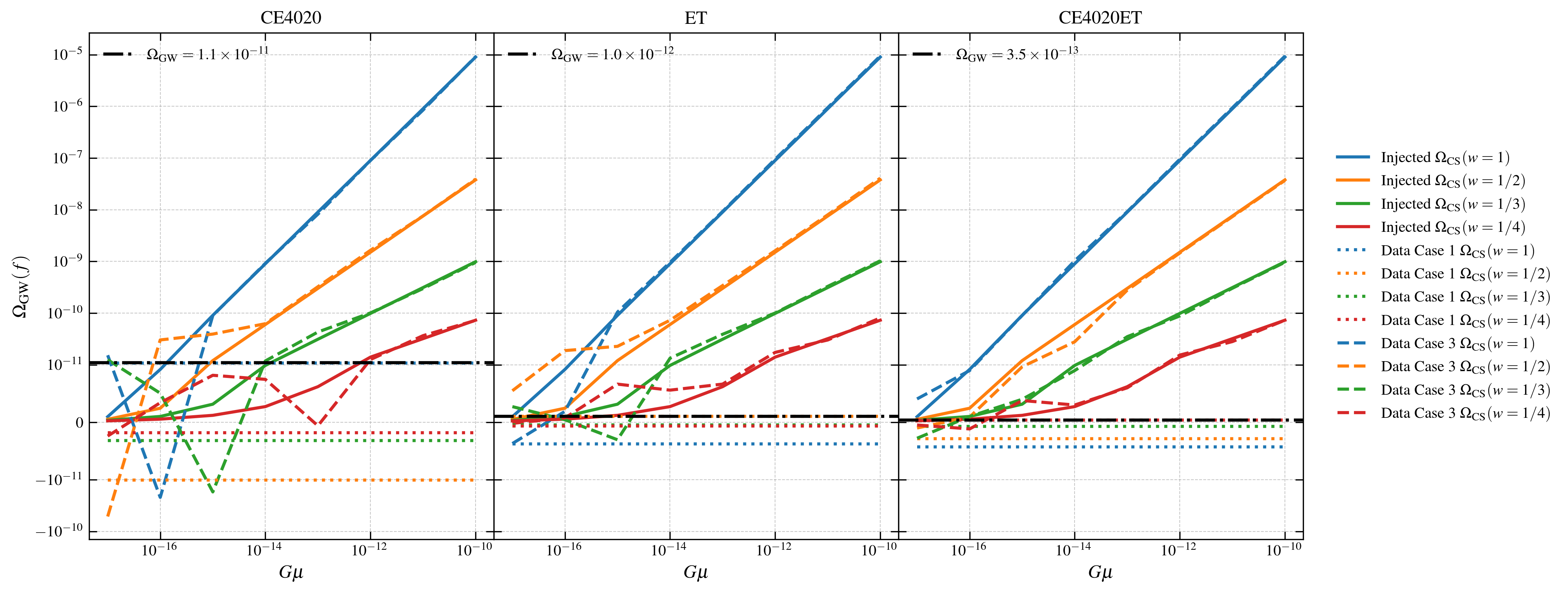}
    \caption{The joint estimation of the CS fractional energy density spectrum with data cases 1 and 3. \textit{Left}: the CE4020 network. \textit{Middle}: the ET detector. \textit{Right}: the CE4020ET network. The dotted line represents $\hat{\Omega}_{\rm CS}$ in data case 1, which includes only foreground noise. The solid line represents the injected $\Omega_{\rm CS}$, while the dashed line represents the joint estimation $\hat{\Omega}_{\rm CS}$ of the CS signal in data case 3.}
    \label{Foregounds}
\end{figure*}

Additionally, Fig.~\ref{Foregounds} presents a comparison of the foreground separation results for data cases 1 and 3 across the three detector networks. The findings indicate that, in the case where only astrophysical foreground noise is present (data case 1), the estimated CS spectrum obtained via the component separation method is independent of both the CS model and $G\mu$, consistently yielding a small value. This result arises from the incomplete separation of foreground components.
Furthermore, as the number of detectors increases, the estimated amplitude of the CS spectrum decreases. For the CE4020 network, the estimated CS spectrum is approximately $1.1 \times 10^{-11}$; for the ET detector, it is around $1.0 \times 10^{-12}$; and for the CE4020ET network, it is roughly $3.5 \times 10^{-13}$. Comparing these results with those from data case 3 allows us to evaluate the foreground separation capabilities of each detector network.

\begin{figure*}
    \centering
    \includegraphics[width=1\linewidth]{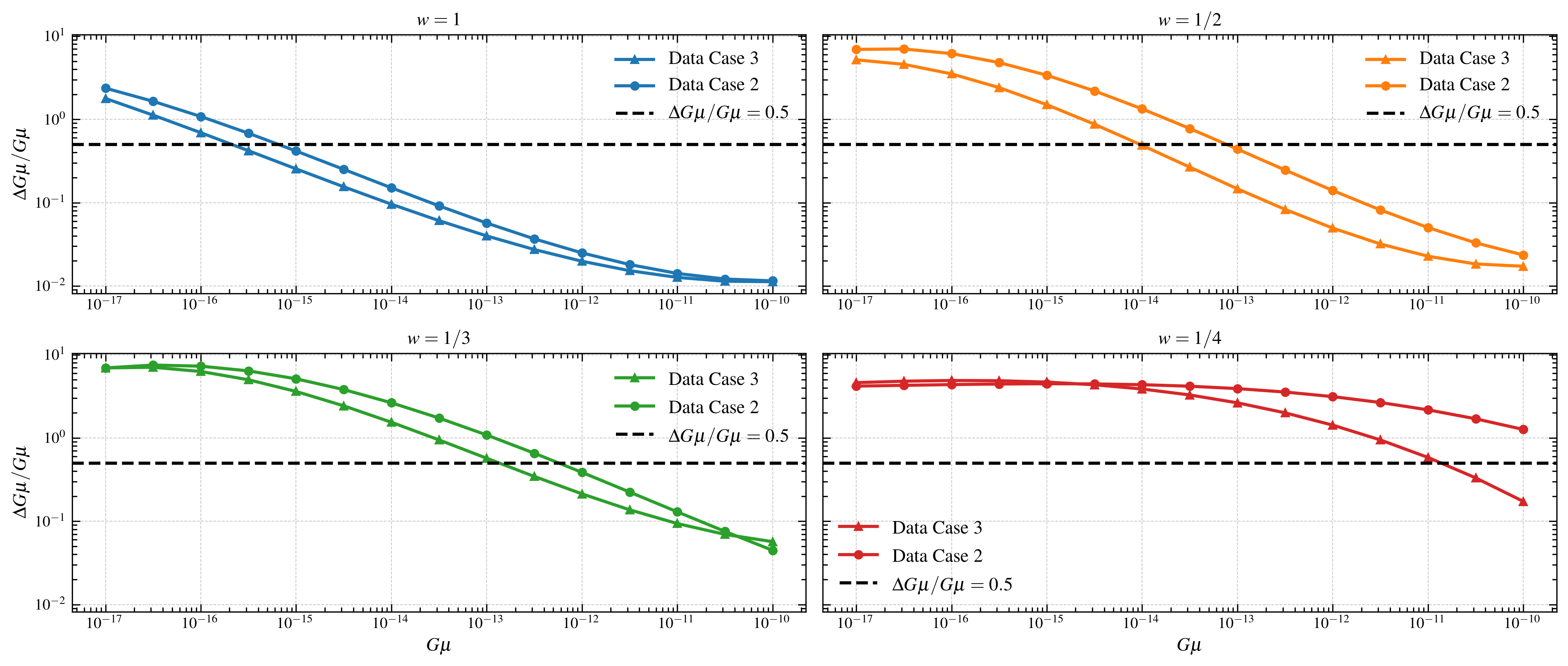}
    \caption{The component separation estimation of the uncertainty of $G\mu$ for CE4020 network under different equation-of-state parameters $w$. We compare scenarios considering only CBC foregrounds with those fully incorporating both CBC and CBHE foregrounds. The line with triangular markers represent data case 3, while the line with circular markers represent data case 2. The black dashed line represents the uncertainty limit $\Delta G\mu / G\mu = 0.5$.}
    \label{DeltaGmu1}
\end{figure*}

\begin{figure*}
    \centering
    \includegraphics[width=1\linewidth]{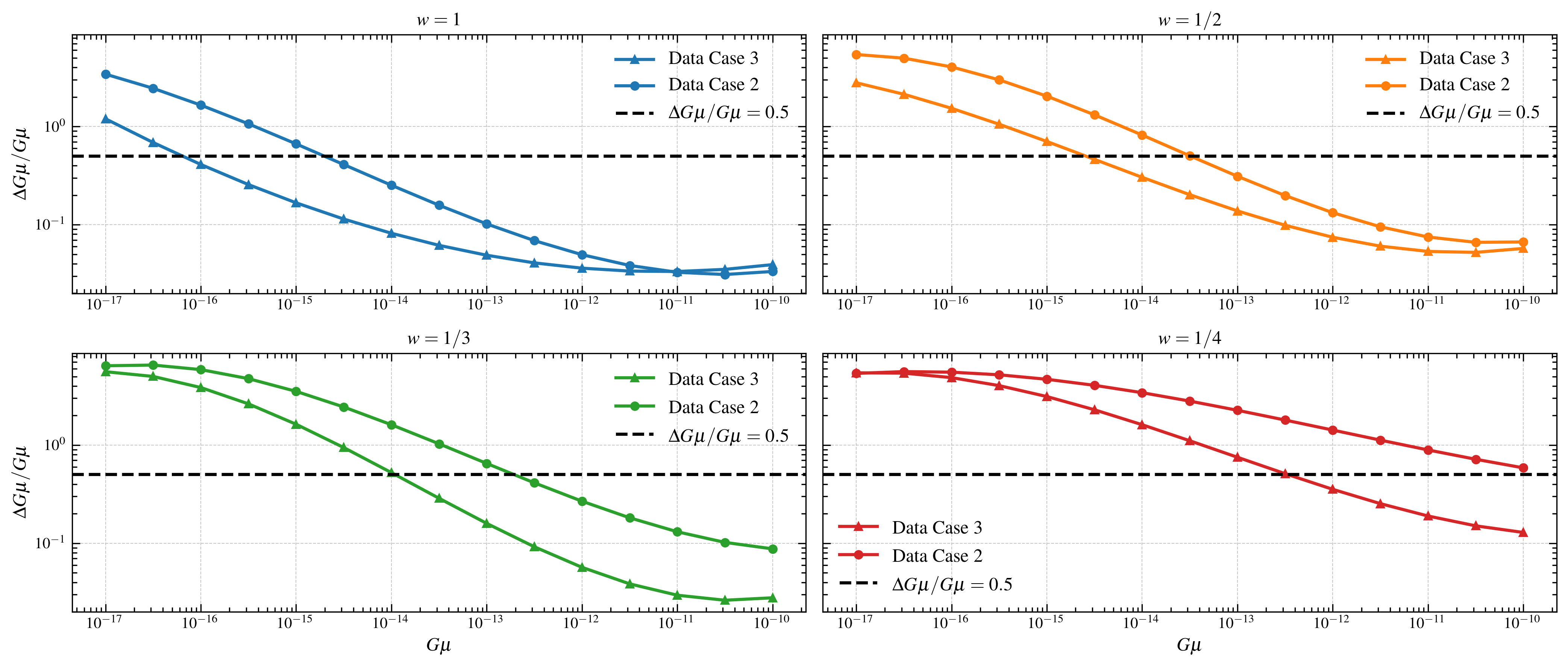}
    \caption{The component separation estimation of the uncertainty of $G\mu$ for ET detector under different equation-of-state parameters $w$. We compare scenarios considering only CBC foregrounds with those fully incorporating both CBC and CBHE foregrounds. The line with triangular markers represent data case 3, while the line with circular markers represent data case 2. The black dashed line represents the uncertainty limit $\Delta G\mu / G\mu = 0.5$.}
    \label{DeltaGmu2}
\end{figure*}

\begin{figure*}
    \centering
    \includegraphics[width=1\linewidth]{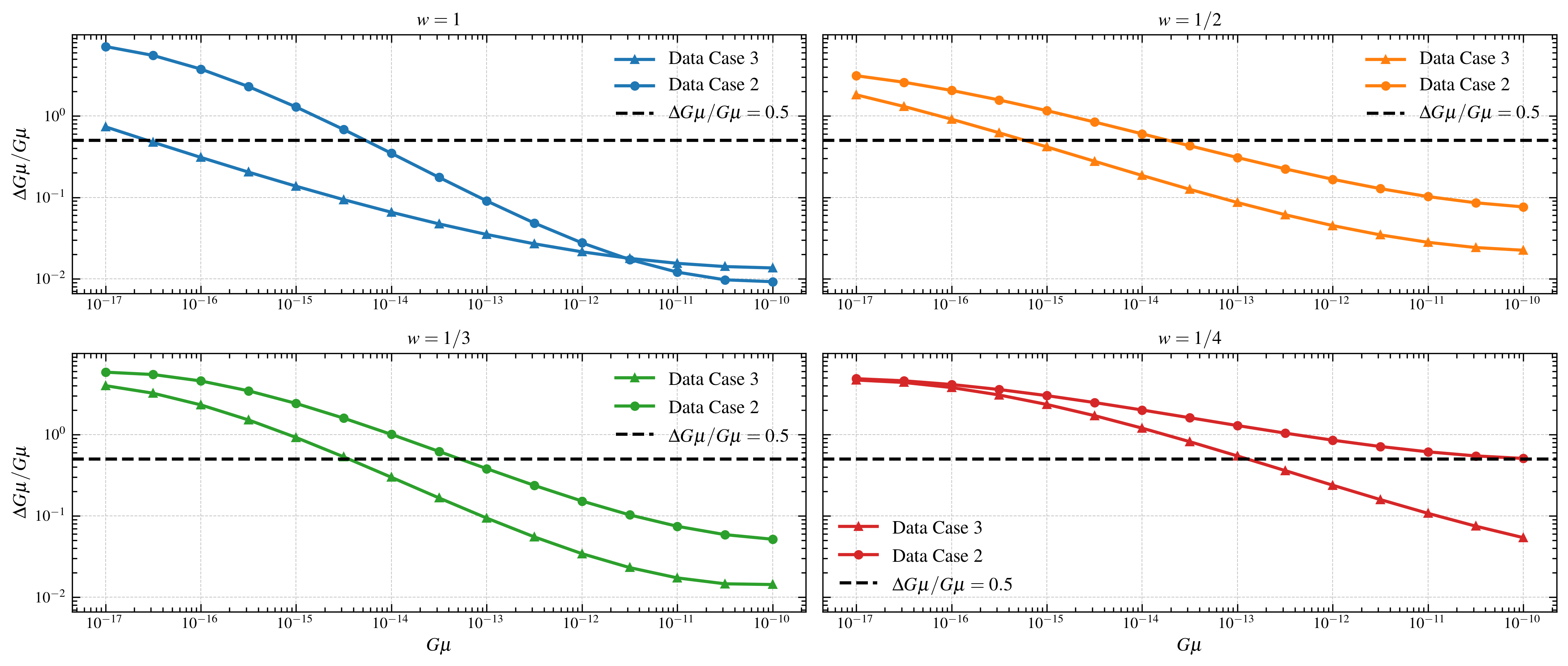}
    \caption{The component separation estimation of the uncertainty of $G\mu$ for CE4020ET network under different equation-of-state parameters $w$. We compare scenarios considering only CBC foregrounds with those fully incorporating both CBC and CBHE foregrounds. The line with triangular markers represent data case 3, while the line with circular markers represent data case 2. The black dashed line represents the uncertainty limit $\Delta G\mu / G\mu = 0.5$.}
    \label{DeltaGmu3}
\end{figure*}

Using the joint estimation of the SGWB spectrum $\hat{\Omega}$ and the covariance $\Sigma$, we apply a Bayesian approach to estimate the uncertainty for $G\mu$. 
\begin{align}
    \log \mathcal{L}(\hat{\Omega} | \theta) = & - \frac{1}{2} \left[ \left( \Omega(\theta) - \hat{\Omega} \right)^{\rm T} \Sigma^{-1} \left( \Omega(\theta) - \hat{\Omega} \right) \right. \nonumber \\
    & \quad + \left. k \log (2 \pi \Sigma) \right],
\end{align}
where $\Omega(\theta)$ represents the SGWB model, $\theta$ denotes its parameters, and $k$ is the number of parameters in the model. We use log-uniform priors in the range [-18, -8] for the three parameters $G\mu$, $\Omega_{\rm CBC}$, and $\Omega_{\rm CBHE}$, and employ the Markov Chain Monte Carlo (MCMC) sampler \texttt{emcee}~\cite{Foreman-Mackey:2012any}.
The uncertainty estimate for $G\mu$ using the component separation method with three networks are shown in Figs.~\ref{DeltaGmu1},~\ref{DeltaGmu2}, and~\ref{DeltaGmu3}. We conduct the study with different SGWBs generated by CSs by changing $G\mu$ and the equation-of-state parameter $w$. We consider that the CS SGWB cannot be separated when $\Delta G\mu / G\mu > 0.5$~\cite{Boileau:2021gbr, Wang:2023ltz}. The black dashed line represents the uncertainty limit $\Delta G\mu / G\mu = 0.5$. The line with triangular markers represent the component separation results with the full consideration of both CBC and CBHE foregrounds (data case 3), while the line with circular markers represent the results without considering the contribution of the CBHE foreground during component separation (data case 2). Table~\ref{tab:my_label} summarizes the capability of three networks to estimate the CS tension. 

\begin{table*}
    \centering
    \begin{tabular}{c|c|c|c|c}
        \hline
         \multirow{2}{*}{} & \multirow{2}{*}{$w$} & \multicolumn{3}{c}{$G\mu$} \\ \cline{3-5}
         & & CE4020 & ET & CE4020ET \\ \hline
        \multirow{4}{*}{Data Case 2} & $1$ & $6.3 \times 10^{-16}$ & $2.0 \times 10^{-15}$ & $5.6 \times 10^{-15}$ \\ 
         & $1/2$ & $5.6 \times 10^{-14}$ & $3.2 \times 10^{-14}$ & $2.0 \times 10^{-14}$ \\
         & $1/3$ & $5.6 \times 10^{-13}$ & $2.0 \times 10^{-13}$ & $5.0 \times 10^{-14}$\\
         & $1/4$ & $>1.0 \times 10^{-10}$ & $>1.0 \times 10^{-10}$ & $>1.0 \times 10^{-10}$ \\ \hline
        \multirow{4}{*}{Data Case 3} & $1$ & $2.2 \times 10^{-16}$ & $6.3 \times 10^{-17}$ & $2.8 \times 10^{-17}$ \\ 
         & $1/2$ & $8.9 \times 10^{-15}$ & $2.5 \times 10^{-15}$ & $5.6 \times 10^{-16}$ \\
         & $1/3$ & $1.4 \times 10^{-13}$ & $1.1 \times 10^{-14}$ & $3.5 \times 10^{-15}$ \\
         & $1/4$ & $1.4 \times 10^{-11}$ & $3.5 \times 10^{-13}$ & $1.3 \times 10^{-13}$ \\ \hline
    \end{tabular}
    \caption{Detection capability of SGWB signals from CSs in three networks. We summarize the results for detecting SGWB signals from CSs under different equation-of-state parameters $w$ across three networks. The comparison includes scenarios considering only CBC foregrounds and those fully incorporating both CBC and CBHE foregrounds. The reported values correspond to cases where the uncertainty in the string tension estimation is $\Delta G \mu / G \mu = 0.5$.}
    \label{tab:my_label}
\end{table*}

The results demonstrate that, for data case 3, the CE4020ET network achieves the highest sensitivity in CS constraints across all CS models. In the standard cosmological scenario ($w = 1/3$), a constraint precision of $G\mu > 3.5 \times 10^{-15}$ can be achieved for $\Delta G \mu / G \mu < 0.5$. For data case 2, comparing data cases 2 and 3 shows that neglecting the CBHE foreground reduces the precision of CS tension constraints. This effect is relatively weak in the CE4020 network, as shown in Fig.~\ref{DeltaGmu1}, but it becomes much more pronounced in the ET detector and the CE4020ET network, as illustrated in Figs.~\ref{DeltaGmu2} and~\ref{DeltaGmu3}. Ignoring the CBHE foreground significantly increases the error and decreases detection sensitivity, potentially reducing the precision of CS tension constraints by more than one order of magnitude.

\section{Conclusion}\label{sec6}
This study explores the potential of NG ground-based GW detector networks (CE4020, ET, and CE4020ET) to detect and estimate the parameters of the SGWB from CSs. We analyzed one year of simulated data to constrain CS models under both standard and non-standard cosmological scenarios. Our analysis includes a wider range of astrophysical foreground noise, incorporating contributions from CBCs and CBHEs. 

Our results show that future joint observations with ground-based GW detector networks can impose strict constraints on $G\mu$, achieving an uncertainty of $\Delta G\mu / G\mu < 0.5$ from astrophysical foreground noise. The CE4020ET network has the best observational capabilities, improving the observational precision by nearly an order of magnitude compared to individual ET and CE detector networks. In the standard cosmological scenario with $w = 1/3$, the CE4020 network and ET detectors can impose precise constraints on CSs for $G\mu > 1.4 \times 10^{-13}$ and $G\mu > 1.1 \times 10^{-14}$. The CE4020ET network provides even better constraints, reaching $G\mu > 3.5 \times 10^{-15}$.

At the same time, our results highlight the critical importance of astrophysical foreground modeling for future GW detector networks. We studied the CBHE foreground and found that failing to properly account for it can degrade the estimation accuracy of CS tension. For the CE4020 network detectors, this effect is relatively small, but for the ET and CE4020ET network, the impact is significant, potentially reducing the accuracy by one to two orders of magnitude, with the effect becoming more pronounced at $w = 1$ and $w = 1/4$. The future joint observations of the SGWB with ET and CE detectors have great potential, and our results indicate that the CBHE foreground contribution cannot be ignored in the parameter inference of high-sensitivity networks.

We also found that the CE4020ET network is expected to identify non-standard features in the SGWB spectrum from CSs, thus helping to uncover the pre-BBN history of the universe. For $w = 1$, the CE4020ET network is capable of placing a constraint on $G\mu$ with an uncertainty of $\Delta G\mu/G\mu < 0.5$ for $G\mu$ values above $2.8 \times 10^{-17}$. In the case of $w = 1/2$, the CE4020ET network can provide a precision on $G\mu$ with an uncertainty of $\Delta G\mu/G\mu < 0.5$ for $G\mu > 5.6 \times 10^{-16}$. For $w = 1/4$, the CE4020ET network can provide a precision on $G\mu$ with an uncertainty of $\Delta G\mu/G\mu < 0.5$ for $G\mu > 1.3 \times 10^{-13}$.

Future research will expand on these results by including additional sources of cosmological SGWB, such as those from first-order phase transitions and inflation. Multi-band GW observations can enhance the accumulation of signal-to-noise ratio and help modeling astrophysical foregrounds~\cite{Sesana:2016ljz, Vitale:2016rfr, Zhang:2021pwe, Yang:2022iwn, Jin:2023zhi, Dong:2024bvw}. We plan to combine future space-based GW detectors LISA~\cite{LISA:2017pwj}, Taiji~\cite{Hu:2017mde}, and TianQin~\cite{TianQin:2015yph} for multi-band joint observations, which are expected to offer more precise constraints on the non-standard CS energy spectrum~\cite{Marriott-Best:2024anh}. These efforts will enhance our understanding of cosmological GW backgrounds and further improve constraints on CS models. 

\section*{Acknowledgments}
This work was supported by the National Natural Science Foundation of China (Grant Nos. 12473001 and 11975072), the National SKA Program of China (Grants Nos. 2022SKA0110200 and 2022SKA0110203), and the National 111 Project (Grant No. B16009).

\bibliography{main}

\end{document}